\documentclass[preprint,showpacs,preprintnumbers,amsmath,amssymb]{revtex4}
\usepackage{graphicx}
\usepackage{dcolumn}
\usepackage{bm}

\begin{document}

\title{Electronic Transport Spectroscopy of Carbon Nanotubes in a Magnetic Field}

\author{P. Jarillo-Herrero, J. Kong, H. S. J. van der Zant, C. Dekker, L. P. Kouwenhoven, and S. De Franceschi}
\affiliation{ Kavli Institute of Nanoscience, Delft University of
Technology, PO Box 5046, 2600 GA Delft, The Netherlands}
\date{\today}
\begin{abstract}
We report magnetic field spectroscopy measurements in carbon
nanotube quantum dots exhibiting four-fold shell structure in the
energy level spectrum. The magnetic field induces a large
splitting between the two orbital states of each shell,
demonstrating their opposite magnetic moment and determining
transitions in the spin and orbital configuration of the quantum
dot ground state. We use inelastic cotunneling spectroscopy to
accurately resolve the spin and orbital contributions to the
magnetic moment. A small coupling is found between orbitals with
opposite magnetic moment leading to anticrossing behavior at zero
field.
\end{abstract}

\pacs{73.22.-f, 73.22.Dj, 73.23.Hk, 73.63.Fg}
\maketitle

The remarkable electronic behavior of carbon nanotubes (CNTs)
originates from a particular combination of the symmetry
properties of the graphene band structure and the quantization of
momentum imposed by periodic boundary conditions along the
nanotube circumference~\cite{Dressel,review}. This symmetry
results in a four-fold degenerate shell structure in the energy
spectrum of CNT quantum dots (QDs). In early experiments on CNT
QDs~\cite{Tans,Bockrath}, however, this symmetry was not observed,
presumably due to the presence of defects. Improvements in the
quality of CNTs and advances in nanofabrication techniques have
enabled the recent observation of four-fold degeneracy in the
spectrum of CNT QDs~\cite{Liang,Buitelaar}. An interesting effect
related to the symmetry of the graphene band structure is the
modulation of the energy gap in CNTs when placed in a parallel
magnetic field, $B$. This effect, predicted early on~\cite{Ando},
has only been recently observed in CNT QDs~\cite{Minot,Coskun} .
These studies, however, did not show evidence of four-fold
symmetry and the link between the energy spectrum and the
$B$-evolution of the QD states was not established.

In this Letter, we report $B$-dependent electronic transport
spectroscopy measurements on CNT QDs exhibiting four-fold shell
structure. We show that: (i) each shell consists of two orbitals
with opposite magnetic moment; (ii) the splitting of the orbital
states with $B$ accounts for all the observed transitions in the
spin and orbital configuration of the CNT QD; (iii) a weak
coupling exists between orbitals with opposite magnetic moment
resulting in level repulsion at $B$=0; (iv) Zeeman and orbital
contributions to the electron magnetic moment can be distinguished
by inelastic-cotunneling spectroscopy.

The electronic structure of CNTs can be derived from the
two-dimensional band structure of graphene. The continuity of the
electron wave function around the CNT circumference imposes the
quantization of the wave-vector component perpendicular to the CNT
axis, $k_{\perp}$. This leads to a set of one-dimensional subbands
in the longitudinal direction~\cite{Dressel}. Due to symmetry, for
a given subband at $k_{\perp}$=-$k_{o}$ there is a second
degenerate subband at $k_{\perp}$=$k_{o}$. Figure 1a shows in
black solid lines the schematic 1D band structure of a gapped CNT
near the energy band gap. Both valence and conduction bands have
two degenerate subbands, labelled as ``+'' and ``$-$''. Ajiki and
Ando~\cite{Ando} predicted that the orbital degeneracy should be
lifted by a magnetic field parallel to the CNT axis (Fig.1a). This
effect can be understood by noting that, due to clockwise and
anti-clockwise motion around the tube, electrons in degenerate
``+'' and ``$-$'' subbands should have opposite orbital magnetic
moments, $\mu_{orb}$. In the case of finite-length CNTs, a
discrete energy spectrum is expected from size quantization. The
level spectrum of a CNT QD can then be described as two sets of
spin-degenerate levels, $E_{+}^{(n)}$ and $E_{-}^{(n)}$ with
$n$=1,2,3,...(see Fig. 1a). In the absence of inter-subband
mixing, $E_{+}^{(n)} = E_{-}^{(n)}$ at $B$=0, and a four-fold
degenerate shell structure is expected for every $n$. (Below we
show that a finite coupling can exist, resulting in a small
orbital splitting even at $B$=0.)

The four-fold shell filling emerges in a measurement of the linear
conductance, $G$, versus gate voltage, $V_ {G}$. This is shown in
Fig. 1b for a QD device fabricated from a metallic nanotube with a
small band gap~\cite{Minot,Notebandgap,NoteFabrication}. $G$
exhibits Coulomb blockade oscillations~\cite{review} corresponding
to the filling of the ``valence'' band of the CNT. From left to
right, electrons are consecutively added to the last three
electronic shells, $n$=3, 2 and 1, respectively. The shell
structure is apparent from the $V_{G}$-spacing between the Coulomb
oscillations. The addition of an electron to a higher shell
requires an extra energy cost corresponding to the energy spacing
between shells. This translates into a larger width of the Coulomb
valley associated with a full shell~\cite{Notelevel}. The first
group of four Coulomb peaks on the left-hand side of Fig. 1b
($n$=3) are strongly overlapped due to a large tunnel coupling to
the leads and Kondo effect~\cite{Kondopub}. The coupling decreases
with $V_{G}$, becoming very small near the band gap, which lies
just beyond the right-hand side of the $V_{G}$-range shown. Due to
this small coupling, the Coulomb peaks associated with the last
two electrons in $n$=1 are not visible (see Fig 2b).

The shell structure breaks up at finite $B$ (Fig. 2a). In each
group of four Coulomb peaks, the first (last) two peaks shift
towards lower (higher) $V_{G}$. This behavior demonstrates the
strong $B$-dependence of the orbital levels described in Fig. 1a.
The magnetic field shifts the ``$-$'' orbital levels down in
energy, while the ``+'' orbitals are shifted up due to their
opposite $\mu_{orb}$. Consequently, the addition of the first
(last) two electrons to a shell results in a pair of Coulomb peaks
shifting toward lower (higher) $V_{G}$. For each shell,
$\mu_{orb}$ can be extracted from the shift, $\Delta V_{G}(n)$, in
the position of the corresponding Coulomb peaks. Neglecting the
Zeeman splitting, we use the relation $e\alpha\Delta V_{G}(n) =
|\mu_{orb}(n)cos\varphi\Delta B|$, where $\Delta B$ is the change
in $B$, $\varphi$ is the angle between the nanotube and $B$, and
$\alpha$ is a capacitance ratio extracted from non-linear
measurements. The values obtained (0.90, 0.80 and 0.88 meV/T, for
$n$=1, 2 and 3, respectively) are an order of magnitude larger
than the electron spin magnetic moment $(1/2g\mu_{B}$=0.058 meV/T
for $g$=2), and in good agreement with an estimate of $\mu_{orb}$
based on the nanotube diameter~\cite{Notediameter}.

The strong $B$-dependence of the orbital states induces changes in
the orbital and spin configuration of the QD. These are reflected
as kinks in the evolution of the Coulomb peaks with $B$ (Fig. 2b).
Remarkably, a fully consistent description of the $B$-dependent
energy spectrum and the ground state spin/orbital configuration
can be obtained through a careful analysis of all the kinks in
Fig. 2b, as illustrated by the diagrams in Fig. 3. As an example,
we examine the non-trivial evolution of Coulomb peak CC' (notation
defined in Fig. 3 caption). Segment $\mathrm{CC_1}$ separates the
triplet state in region II from the spin 1/2 state in region III.
The incoming electron tunnels into the $E_{-}^{(2)}$ orbital, with
spin down, so the slope of the $\mathrm{CC_1}$ segment is
``$-\mu_{orb}^{(2)} + 1/2g\mu_{B}$'', as noted underneath the
corresponding arrow. At $\mathrm{B_1C_1}$, a triplet-singlet
transition occurs. Therefore $\mathrm{C_1C_2}$ separates the
singlet state in region II from the spin 1/2 state in region III.
Now the incoming electron tunnels into the $E_{+}^{(2)}$ orbital
state with spin up, so the slope of $\mathrm{C_1C_2}$ is
``$\mu_{orb}^{(2)} - 1/2g\mu_{B}$''. Interestingly at
$\mathrm{C_2}$, a kink related to an inter-shell orbital crossing
occurs. $\mathrm{C_2C'}$ also separates the singlet state in
region II from the spin 1/2 state in region III (as
$\mathrm{C_1C_2}$), but the incoming electron tunnels into the
$E_{-}^{(1)}$ state and with spin up, so the slope changes
direction and has a value ``$-\mu_{orb}^{(1)} - 1/2g\mu_{B}$''.
The rest of the diagrams can be followed in a similar manner.

Note that kinks in Fig. 2b are connected by conductance ridges
crossing the Coulomb valleys. The enhancement of $G$ at these
ridges is due to Kondo effects of different origins. At
$\mathrm{B_1C_1}$, $\mathrm{D_1E_1}$, and $\mathrm{F_1G_1}$ the
Kondo effect arises from singlet-triplet degeneracy~\cite{Sasaki}.
At AB, CD, and EF an enhanced Kondo effect is observed in relation
to orbital degeneracy~\cite{Kondopub}. The Kondo ridges at
$\mathrm{C_2D_2}$ and $\mathrm{E_2F_2}$ are due to the recovery of
orbital degeneracy between $E_{-}^{(2)}$ and
$E_{+}^{(1)}$~\cite{Kondopub}. Note that, as a result of
electron-hole symmetry, region III (three electrons in shell
$n$=2) and region V (one electron in shell $n$=1) have a certain
degree of mirror symmetry, both in terms of the slope of the
Coulomb peaks' evolution with $B$ and the Kondo ridges.

The data shown so far have been explained in terms of a
$B$-induced splitting of orbital degeneracy, as if the two orbital
states of every shell were indeed degenerate at $B$=0.  A small
zero-field orbital splitting may in fact exist and be masked by
the Kondo effect at AB, CD, and EF.  In order to investigate this
possibility, we considered a different device with a much smaller
coupling to the leads and hence much weaker Kondo effect. This
device also exhibits four-fold periodicity in the Coulomb peaks'
pattern. Fig 4c shows a Coulomb diamond corresponding to one
electron in a shell at $B$=80mT~\cite{Superconduc}. Inside the
diamond, single electron tunneling is suppressed and transport
occurs via higher order cotunneling processes. A sharp increase in
the differential conductance, $dI/dV$, is observed at a bias $|V|
= V_{in}\sim190\mu$V, denoting the onset of inelastic cotunneling
(IC)~\cite{Grabert, Silvano, Kogan}. The IC transition takes place
between the two spin-degenerate orbital levels of the same shell
thereby indicating the existence of a finite splitting at $B$=0.
Before discussing the $B$-dependence of the IC edges we note that
a weak Kondo peak is also present at $V$=0 (top inset in Fig. 4a).
This Kondo effect arises from the single-electron occupancy of the
spin degenerate orbital ground state.

At finite $B$, both the Kondo peak and the IC edges split due to
Zeeman spin splitting. This is shown in Fig. 4a, where $dI/dV$ is
plotted vs ($V$,$B$) for $V_{G}$ at the center of the Coulomb
diamond~\cite{Superconduc}. In order to identify the $dI/dV$ steps
more clearly, Fig. 4b shows the numerical derivative of the
$dI/dV$ plot in Fig. 4a (i.e. $d^{2}I/dV^{2}$ vs $V$ and $B$). IC
steps in Fig. 4a turn into peaks ($V$$>$0) or dips ($V$$<$0) in
Fig. 4b. The zero-bias Kondo peak evolves into two $dI/dV$ steps
at $V= ±g\mu_{B}B/e$ ($g$ = 2). These correspond to IC processes
in which the spin state of the QD is flipped, i.e. from
$|-,\uparrow>$ (ground state) to $|-,\downarrow>$ (excited state).
Each of the two $dI/dV$ steps associated with inter-orbital IC
splits by $g\mu_{B}B/e$ and they move further apart due the
increasing orbital splitting, ~$2\mu_{orb}Bcos\varphi$
($\varphi=33^\circ$). We estimate 2$\mu_{orb}\sim350\mu e$V/T i.e.
$\sim3$ times the Zeeman splitting~\cite{Notemoment}. The inner
inter-orbital steps correspond to IC from $|-,\uparrow>$ to
$|+,\uparrow>$, and evolve with a slope $\pm2\mu_{orb}/e$. The
outer inter-orbital steps correspond to IC from $|-,\uparrow>$ to
$|+,\downarrow>$, and evolve with a slope $\pm(2\mu_{orb} +
g\mu_{B}B)/e$. The six steps ("Zeeman", "orbital" and
"orbital+Zeeman") can be seen in the bottom inset to Fig. 4a. Such
separation between the orbital and Zeeman contributions to the
magnetic moment of electrons in CNTs has not been shown before.
The evolution of the outer IC peaks is non-linear at low $B$,
indicating an anti-crossing between the ``+'' and ``$-$'' orbital
levels. Such IC spectrum can be readily modelled using a
Hamiltonian that includes an inter-orbital coupling besides
orbital and Zeeman terms. The corresponding energy eigenstates
are: $E = \pm\sqrt{(\delta/2)^2+(\mu_{orb}B)^2}\pm 1/2g\mu_{B}B$
(the 4 possible sign combinations). The IC spectrum calculated
with this simple model (Fig. 4d) clearly accounts for the
experimental data. The non-linear evolution of the orbital
splitting with $B$ constitutes direct evidence that the so-called
"subband level mismatch", usually denoted by
$\delta$~\cite{Liang}, is due to a small, but finite, quantum
mechanical coupling between the two orbital subbands in carbon
nanotubes.

We finally comment on the reproducibility of the results shown. $G
(B,V_G)$ patterns with a strong orbital contribution to the
magnetic moment of electrons (similar to that in Fig. 2b) have
been measured in all devices (five from three different
fabrication runs) where four-fold shell filling was
observed~\cite{Noterepro}. Our study demonstrates that the spin
and orbital configuration of CNT QDs can be understood and
controlled by means of a magnetic field. By varying $B$ and the
angle nanotube-$B$, researchers have two ``semi-independent''
knobs ($B$ controls the Zeeman splitting and the angle
nanotube-$B$ controls the orbital splitting), which will prove
very useful in a variety of experiments with CNT QDs, such as the
study of the Kondo effect in degenerate systems or the interaction
between orbital states at high $B$.

Financial support is obtained from the Dutch organization for
Fundamental Research on Matter (FOM).

\begin{figure}[ht!]
\includegraphics[angle=0,width=0.75\linewidth]{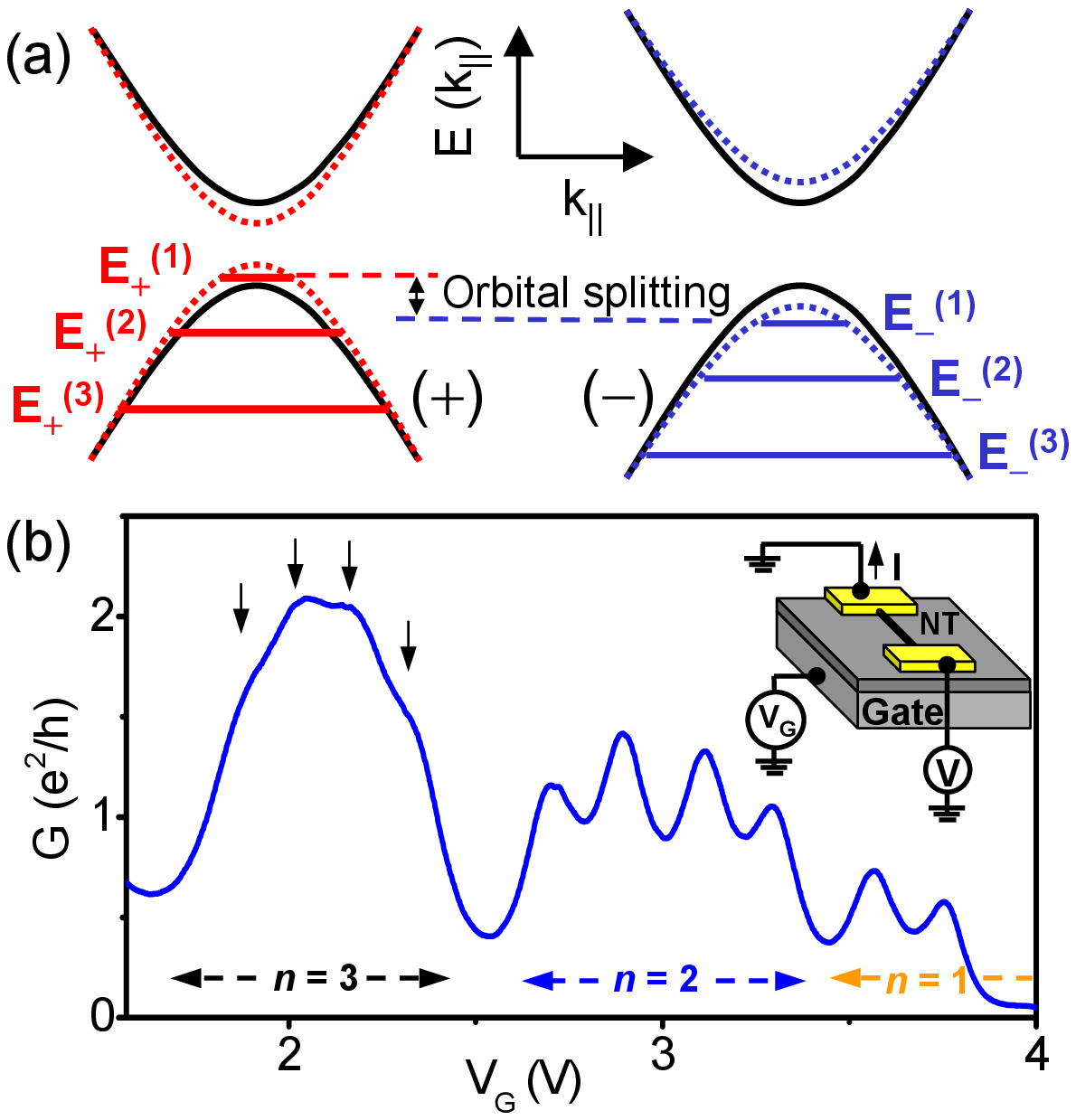}
\caption{\label{fig1} (a) Schematic band structure of a CNT near
its energy gap. Black lines represent the one-dimensional energy
dispersion relation, $E(k_{||})$, at $B$=0 ($k_{||}$ is the wave
vector along the CNT axis). The valence (conduction) band has two
degenerate maxima (minima). Size quantization in a finite-length
CNT results in a set of discrete levels with both spin and orbital
degeneracy. The degeneracy is lifted by a magnetic field parallel
to the CNT. The 1D subbands (and the corresponding levels) at
finite $B$ are represented by red and blue dotted (solid) lines.
Only the orbital splitting of the energy levels is shown in this
figure. (b) Linear conductance, $G$, vs gate voltage, $V_G$ taken
at $T$=8K. Inset: Device scheme.}
\end{figure}

\begin{figure}[ht!]
\includegraphics[angle=0,width=1.0\linewidth]{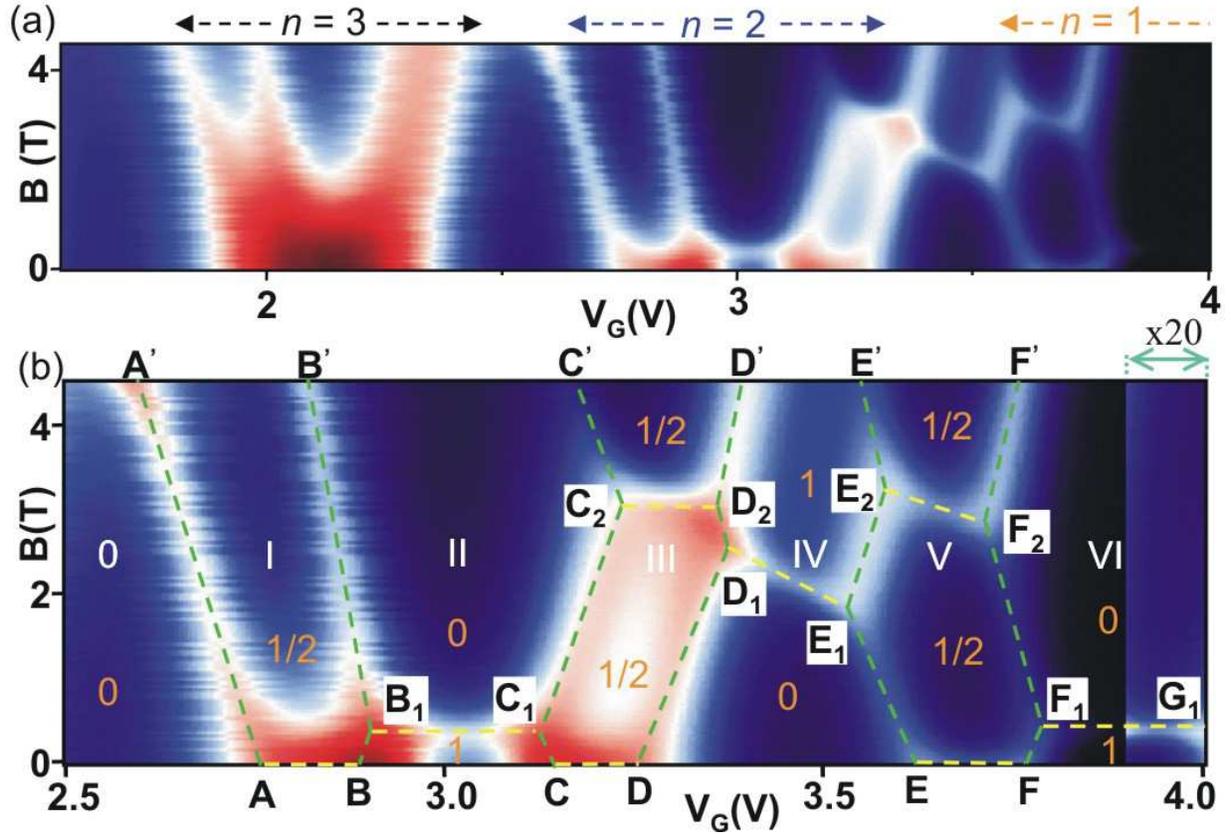}
\caption{\label{fig2} (a) $G$ vs $B$ on a color scale at $T$=0.34K
for the $V_G$ range shown in Fig. 1b (red indicates high $G$ and
dark blue low $G$). (b) Zoom-in of (a). The green dashed lines
highlight the evolution of the Coulomb peaks with $B$. They are
labelled as AA', BB'... and FF'. These divide the plot in
different Coulomb blockade regions indicated by the number of
electrons in the last two shells (white numbers 0 to VI). The
high-$G$ regions (indicated by yellow dashed lines) in between
Coulomb peaks are due to Kondo effect. Orange numbers indicate the
spin in each region. On the right side, the $G$ is multiplied by
20, so that the triplet-singlet transition is clearly seen along
$\mathrm{F_1G_1}$.}
\end{figure}

\begin{figure}[ht!]
\includegraphics[angle=270,width=1.0\linewidth]{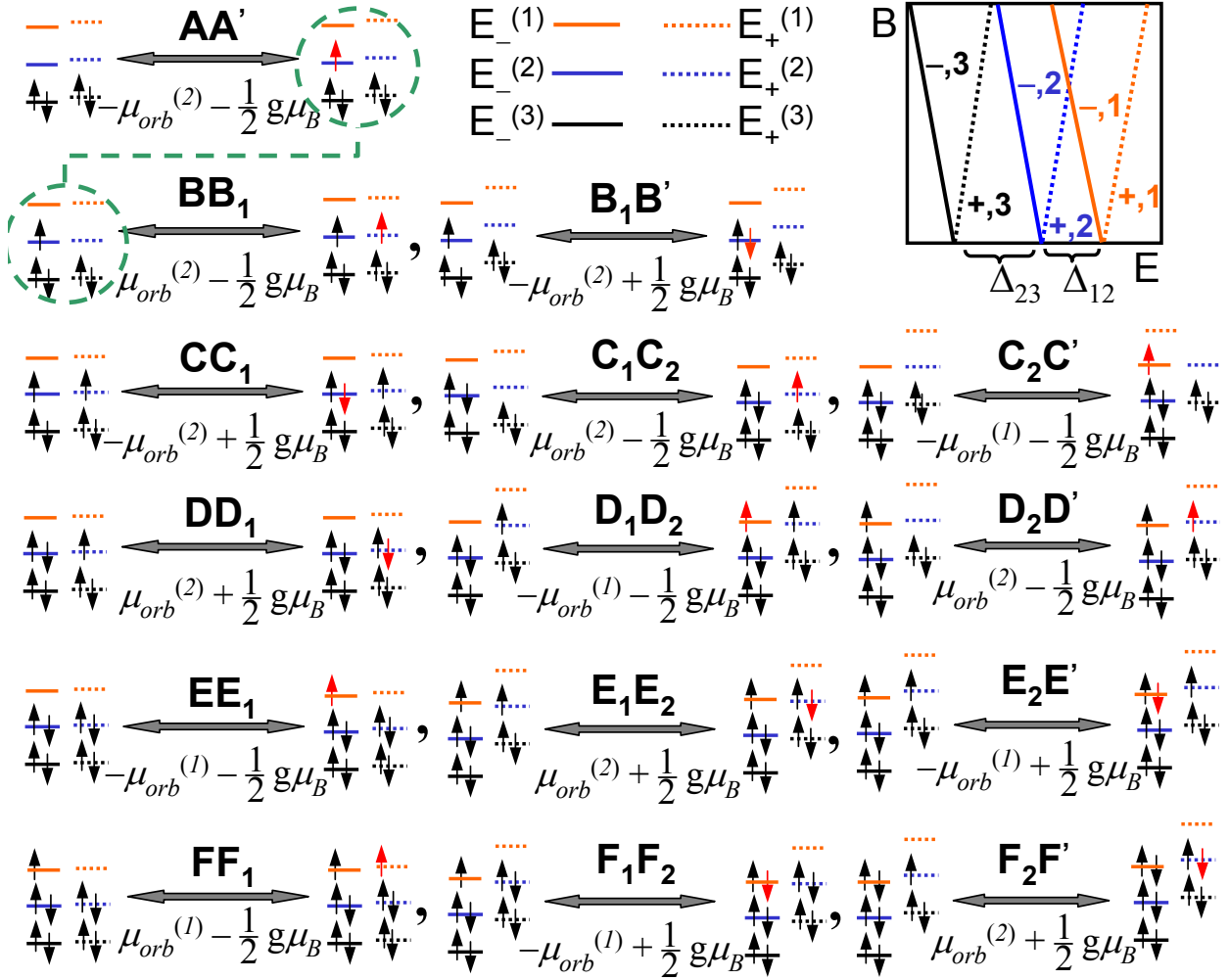}
\caption{\label{fig3}Diagrams representing the orbital and spin
configuration in the different regions of Fig. 2b. Each row
follows the $B$-evolution of a given Coulomb peak. Each section in
a given row shows two diagrams (separated by a double arrow).
These represent the ground state configuration in the two regions
separated by the corresponding segment of the Coulomb peak
evolution with $B$. In the diagrams, different colors refer to
different shells: orange, blue and black for $n$=1, 2 and 3,
respectively (in our discussion here $n$=3 is always filled). The
last added electron is displayed in red, all the others in black.
Note that, for each diagram, the final state is the same as the
initial state for the diagram immediately below (as indicated by
the connected green dashed circles). We use solid (dotted) lines
to represent levels with positive (negative) $\mu_{orb}$, shifting
down (up) with $B$ (Zeeman splitting is neglected because it is an
order of magnitude smaller than the orbital splitting). The slope
corresponding to the $B$-evolution of the Coulomb peaks is also
indicated under the double arrow. Top-right inset: qualitative
energy spectrum of the CNT QD as a function of $B$ (Zeeman
splitting neglected).}
\end{figure}

\begin{figure}[ht!]
\includegraphics[angle=0,width=1.0\linewidth]{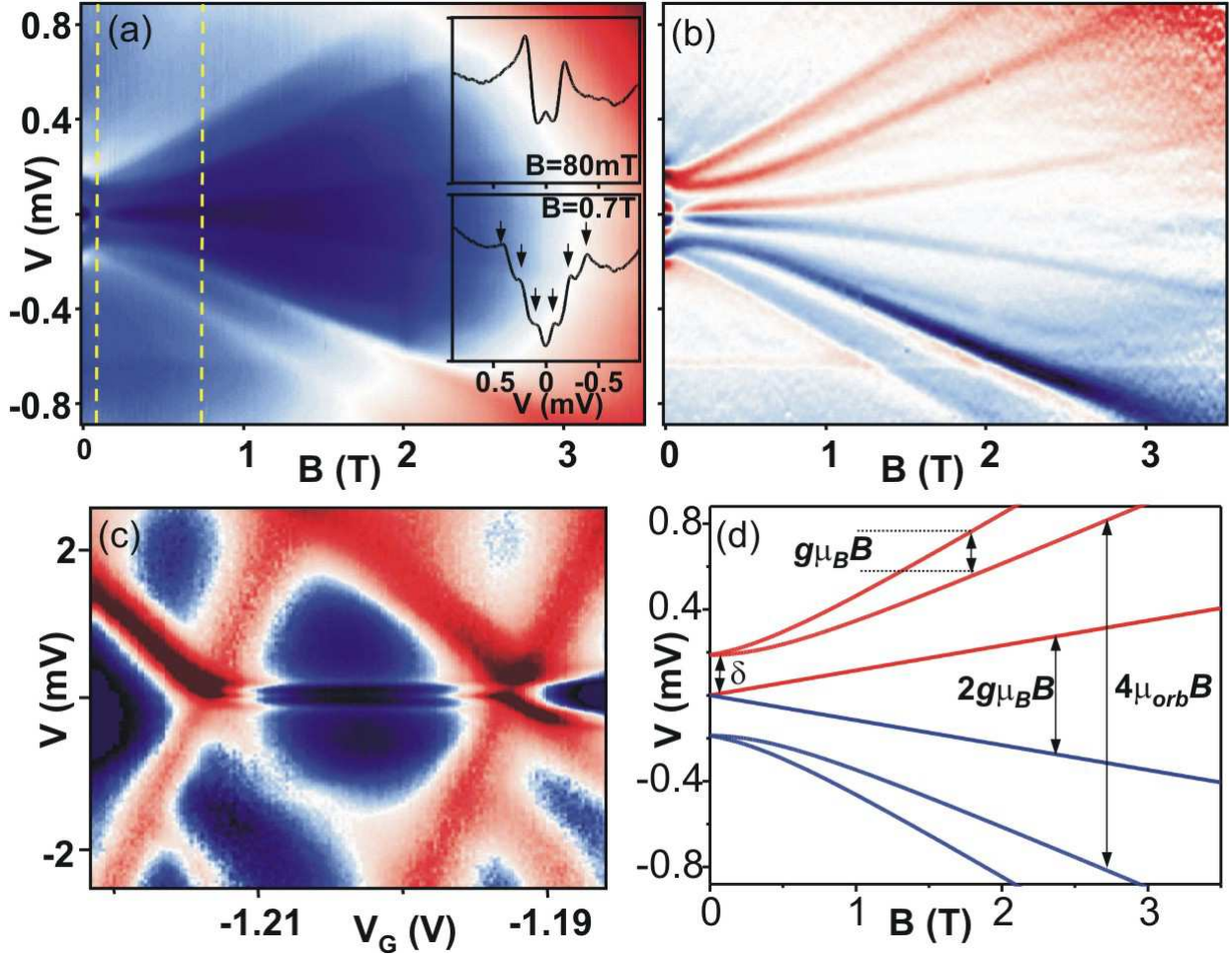}
\caption{\label{fig4}(a) Color-plot of the differential
conductance, $dI/dV$, vs bias, $V$, and $B$, measured in the
center of the Coulomb diamond (see (c)) at $T$=30mK. The yellow
dashed lines indicate the traces shown in the top and bottom
insets. Insets: (top) $dI/dV$ trace taken at $B$=80mT, showing the
onset of inter-orbital IC and a small zero bias peak due to
ordinary spin 1/2 Kondo effect. The vertical axis scale spans from
0.02 to 0.08 $e^2/h$. (bottom) Same as top inset, but at $B$=0.7T,
showing the six IC steps. (b) Numerical derivative of the $dI/dV$
plot in (a). The two inner lines result from Zeeman splitting of
the Kondo peak at $B$=0. The outer lines represent the
$B$-evolution of the spin-split orbital levels. (c) $dI/dV$ vs $V$
and $V_G$, for a single electron in a shell at $B$=80mT. (d)
Calculated $B$-dependence of the  IC spectrum for a single
electron in a spin degenerate level for two coupled orbitals. Red
(blue) lines indicate upwards (downwards) steps in $dI/dV$ with
increasing $V$.}
\end{figure}


\begin{thebibliography}{orbital}
\bibitem{Notebandgap}This band gap can be due to different perturbations, such as curvature or
strain. The measured value of the band gap is $\sim30$meV.
\bibitem{Notelevel}From measurements in the non-linear regime (not
shown), we estimate $\Delta_{1,2}\sim 3$meV, and $\Delta_{2,3}\sim
5$meV, where $\Delta_{n,n+1}\equiv
E_{+}^{(n)}-E_{+}^{(n+1)}=E_{-}^{(n)}-E_{-}^{(n+1)}$. The Coulomb
charging energy is $U\sim 5$meV.
\bibitem{Notediameter} From the value for $n=1$, we extract a
nanotube diameter $D=4\mu_{orb}/ev_F=4.5$nm, in agreement with the
measured diameter of $4.0\pm0.5$nm as determined by atomic force
microscopy. For this device $\varphi=37^\circ$.
\bibitem{Superconduc} At $B<80$mT, the Ti/Au electrodes become
superconducting. The features related to the superconductivity
 in the contacts (see Figs. 4a,b) will be reported elsewhere.
\bibitem{Notemoment} For this device, the measurements are not
taken next to the small-band gap, and this is possibly the reason
why the value of $\mu_{orb}$ is smaller than expected from the
nanotube diameter (2.7nm)~\cite{Minot} (see
also~\cite{Noterepro}).
\bibitem{Noterepro} Still, in some $V_G$ ranges, the
patterns exhibit large variations in $\mu_{orb}$, even for
consecutive shells. This might be due to defects, which could
distort the band structure enough to render the simple picture in
Fig. 1a invalid.
\end{thebibliography}
\end{document}